\documentclass[a4paper,fleqn]{cas-dc}

\usepackage[numbers]{natbib}
\usepackage{indentfirst}
\usepackage{amsmath}

\usepackage{multicol}

\usepackage{cuted} % Provides strip environment for wide equations
\usepackage{stfloats} % Package to control the float placement
\usepackage{lipsum} % For dummy text
\usepackage{upgreek} % For upright Greek letters
\usepackage{titlesec}
\titleformat{\section}{\normalfont\Large\bfseries}{\thesection}{1em}{}

\begin{document}
\let\WriteBookmarks\relax
\def\floatpagepagefraction{1}
\def\textpagefraction{.001}
\let\printorcid\relax

\shorttitle{<HC$^3$L-Diff>}    

\shortauthors{<Shi Yin, Hongqi Tan, et al. >}  

% \title [mode = title]{<main title>}  
\title [mode = title]{HC$^3$L-Diff: Hybrid conditional latent diffusion with high frequency enhancement for CBCT-to-CT synthesis} 

% \author[<aff no>]{<author name>}[<options>]
\author[1,2]{Shi\ Yin}
\fnmark[1]
\author[3,4,5]{Hongqi\ Tan}
\fnmark[1]
\author[1,6,7]{Li\ Ming\ Chong}
\author[1]{Haofeng\ Liu}
\author[2]{Hui\ Liu}
\author[3]{Kang\ Hao\ Lee}
\author[3,4]{Jeffrey\ Kit\ Loong\ Tuan}
\author[1,6,7,8,9,10]{Dean\ Ho}
\author[1,11,7]{Yueming\ Jin\corref{cor1}}
\address[1]{Department of Biomedical Engineering, College of Design and Engineering, National University of Singapore, Singapore}
\address[2]{Institute of Artificial Intelligence \& Robotics (IAIR), Key Laboratory of Traffic Safety on Track of Ministry of Education, School of Traffic and Transportation Engineering, Central South University, Hunan, China}
\address[3]{Division of Radiation Oncology, National Cancer Centre Singapore, Singapore}
\address[4]{Duke-NUS Medical School, Singapore} 
\address[5]{School of Physical and Mathematical Science, Nanyang Technological University, Singapore} 
\address[6]{The Institute for Digital Medicine (WisDM), National University of Singapore, Singapore}
\address[7]{The N.1 Institute for Health (N.1), National University of Singapore, Singapore}
\address[8]{Department of Pharmacology, Yong Loo Lin School of Medicine, National University of Singapore,Singapore}
\address[9]{The Bia-Echo Asia Centre for Reproductive Longevity and Equality (ACRLE), National University of Singapore, Singapore}
\address[10]{Singapore Gastric Cancer Consortium, Department of Medicine, National University of Singapore,  Singapore} 
\address[11]{Department of Electrical and Computer Engineering, College of Design and Engineering, National University of Singapore, Singapore} 

\fntext[fn1]{First Author and Second Author contributed equally to this work.}

\cortext[cor1]{Corresponding author (E-mail: ymjin@nus.edu.sg)}
% Email id of the author
% \ead{ymjin@nus.edu.sg}

\begin{abstract}
\noindent \textit{Background:} Cone-beam computed tomography (CBCT) plays a crucial role in image-guided radiotherapy, but artifacts and noise make them unsuitable for accurate dose calculation. Artificial intelligence methods have shown promise in enhancing CBCT quality to produce synthetic CT (sCT) images. However, existing methods either produce images of suboptimal quality or incur excessive time costs, failing to satisfy clinical practice standards. 

\noindent \textit{Methods and materials:} We propose a novel hybrid conditional latent diffusion model for efficient and accurate CBCT-to-CT synthesis, named HC$^3$L-Diff. We employ the Unified Feature Encoder (UFE) to compress images into a low-dimensional latent space, thereby optimizing computational efficiency. Beyond the use of CBCT images, we propose integrating its high-frequency knowledge as a hybrid condition to guide the diffusion model in generating sCT images with preserved structural details. This high-frequency information is captured using our designed High-Frequency Extractor (HFE). During inference, we utilize denoising diffusion implicit model to facilitate rapid sampling. We construct a new in-house prostate dataset with paired CBCT and CT to validate the effectiveness of our method.

\noindent \textit{Result:} Extensive experimental results demonstrate that our approach outperforms state-of-the-art methods in terms of sCT quality and generation efficiency. Moreover, our medical physicist conducts the dosimetric evaluations to validate the benefit of our method in practical dose calculation, achieving a remarkable 93.8\% gamma passing rate with a 2\%/2mm criterion, superior to other methods.

\noindent \textit{Conclusion:} The proposed HC$^3$L-Diff can efficiently achieve high-quality CBCT-to-CT synthesis in only over 2 mins per patient. Its promising performance in dose calculation shows great potential for enhancing real-world adaptive radiotherapy.
\end{abstract}

% Keywords
% Each keyword is separated by \sep
\begin{keywords}
\sep CBCT-to-CT synthesis
\sep Medical image generation
\sep Latent diffusion model
\sep Dose calculation
\sep Adaptive radiotherapy
\end{keywords}

% \begin{document}
% \begin{frontmatter}
\maketitle

% Main text
\section{Introduction}
Radiotherapy is an integral part of cancer treatment and it uses radiation either in the form of X-ray or energetic particles to deposit dose in the tumor to achieve local control \cite{evans2018principles}. However, due to the nature of the dose deposition curve of photon and particle beam, radiation toxicity is an inevitable consequence of radiotherapy \cite{wang2021radiation}. The main maxim of radiotherapy is therefore to maximise the tumor control while minimizing the normal tissue complication probability. Achieving this balance relies heavily on accurate dose calculation, which is essential for effective and safe treatment delivery.

There is an increasing number of proton therapy facilities being established globally due to the Bragg peak characteristic of the proton dose deposition curve which improves the dose conformity compared to conventional radiotherapy \cite{mohan2022review}. There is clinical evidence suggests a correlation between more conformal dose distributions and reduced side effects. This has been observed in head and neck cancers and the treatment of central nervous system (CNS) tumors \cite{alterio2019modern}. Despite the dosimetric advantage of proton therapy, the integrity of the dose distribution is more susceptible to anatomical changes \cite{pham2022magnetic}. Advancement in image-guided radiotherapy (IGRT) especially the development of cone-beam computed tomography (CBCT) is a cost-effective method to acquire 3D volumetric information of the daily patient's anatomy \cite{franzone2016image}. However, the major drawback of CBCT is the low image quality due to scatters and motions compared with CT, which result in inaccurate Hounsfield units (HU) and make the images unsuitable for dose calculation \cite{kong2016cone}. The ability to calculate doses on the daily CBCT empowers the clinic to make informed clinical decisions based on the individual target and organ-at-risk (OAR) doses. These decisions could be to continue treatment or to arrange for a new CT scan and a replan because the target coverage has been compromised; all of these are essential in moving towards adaptive radiotherapy (ART) \cite{Albertini2020}. 
In this regard, improving the quality of CBCT to the level of CT is crucial and highly demanded to benefit ART workflow. 

Some approaches have been proposed in the literature to enhance CBCT image, including early physics based-method \cite{Jin2010} and Hounsfield look-up table (HLUT) \cite{Lechner2023}. More recently, artificial intelligence (AI) and deep learning methods are emerging and showing great potential to fundamentally change the radiation oncology workflow and pave the path towards online ART, which allows for rapid adaptation of treatment plans to accommodate daily anatomical changes, ensuring consistent target coverage and sparing of surrounding healthy tissues \cite{mccomas2023online, dou2022deep, tortora2021deep}. These AI methods have shown promising results in improving the quality of CBCT to the level of CT. The task is commonly named as CBCT-to-CT synthesis with the improved CBCT known as synthetic CT (sCT) \cite{spadea2021deep}. 
%has recently yielded excellent results at several treatment sites and with different architectures . 
For instance, convolutional neural networks (CNN) \cite{chen2020synthetic} have been used for sCT generation, and have benefitted ART \cite{wang2022synthetic}. Next, generative adversarial networks (GANs) have gradually played an important role in sCT generation. Consisting of a generator and a discriminator, GANs can be optimized through adversarial training, resulting in higher-quality image synthesis compared to traditional CNNs. Several studies have utilized GAN-based methods to improve CBCT-to-CT synthesis \cite{deng2023synthetic, pang2023comparison}. However, training GANs is often unstable and optimization can be challenging, leading to issues such as model collapse \cite{boulanger2021deep}. 

Recently, diffusion model has revolutionized the field of image synthesis \cite{dhariwal2021diffusion}, image editing \cite{Yang_2023_CVPR}, super-resolution \cite{yue2024resshift}, and there are also some literatures that utilize diffusion model for CBCT-to-CT synthesis, showing state-of-the-art results \cite{fu2024energy}. Diffusion model was originally devised for simulating substance diffusion in space \cite{sohl2015deep}, and the advent of denoising diffusion probabilistic models (DDPM) \cite{ho2020denoising} brings the diffusion model to the field of image synthesis by adding noise followed by progressive denoising. Diffusion model involves a process of adding noise to images and then iteratively denoising them based on conditional probability distributions, effectively reconstructs images by sampling from the model's output distributions. Compared to the GAN-based method, it possesses progressive refinement through denoising, improving the detail and clarity of the image \cite{dhariwal2021diffusion}.
For example, Peng et al. \cite{peng2024cbct} use a conditional DDPM to generate sCT images based on CBCT images. The CBCT images are concatenated with the noise samples along the channel dimension as the condition, which is used to guide each sampling generation step of CT, and ensures that the generated sCT images are accurate counterparts of the input CBCT images, effectively enhancing the overall quality of the generated images. Fu et al. \cite{fu2024energy} also utilize the diffusion model and design an energy-guided function that differentiates between domain-independent and domain-specific features, enhancing sCT generation by preserving relevant details and discarding less critical ones.

Although these pilot studies using diffusion model further improve the quality of sCT images, we discern two principal limitations within these literatures, including low efficiency and detail deficiency in sCT generation. (i) The iterative nature of diffusion model-based image generation incurs significant computational time, often exceeding an hour to produce a 3D medical volume of a patient, which does not meet the needs of clinicians, as the maximum inference time to produce sCT for one patient should be less than 5 minutes. 
This strict time constraint is crucial for online ART, ensuring rapid adaptation and timely treatment when substantial anatomical changes require a new treatment plan, minimizing patient discomfort and optimizing outcomes
\cite{green2019practical}.
Recently, the latent diffusion model has emerged and gained a lot of attention \cite{rombach2022high}. Different from the original diffusion models, the latent diffusion model uses autoencoders to compress original images into a lower dimensional latent space, enabling more efficient model training and image generation. It has been applied to various domains in general computer vision of image generation and editing \cite{rombach2022high,wu2023latent}. Also in medical domain, it has shown promise in MRI image synthesis \cite{jiang2023cola}. These achievements inspired us to explore the potential of latent diffusion model in CBCT-to-CT generation.
(ii) The images generated by existing models often exhibit structural ambiguities or inconsistencies, which fail to reflect the complex anatomical structures in CT imaging. The underlying reason may be that current models are primarily conditioned on the original CBCT images, which are insufficient for describing detailed anatomical structures. How to wisely extract and incorporate other informative modalities to augment diffusion process is crucial to improve sCT generation. For example, high-frequency features, can offer additional fine structural information like edges and textures, which can improve the representation of intricate anatomical details.

In this paper, we propose a novel \textbf{\textit{H}}ybrid \textbf{\textit{C}}onditional \textbf{\textit{L}}atent \textbf{\textit{Diff}}usion model for efficient and accurate \textbf{\textit{C}}BCT-to-\textbf{\textit{C}}T image synthesis, named \textbf{\textit{HC$^3$L-Diff}}. Specifically, we first propose the unified feature encoder (UFE) to compress images into a lower-dimensional yet perceptually equivalent latent embedding space. This compression can reduce the training time, computational resources required for the diffusion model, and meanwhile, accelerate the generation process. Moreover, we propose to include both CBCT and its corresponding high-frequency information as a hybrid condition, to guide the generation of sCT images in the latent space. We design a high-frequency extractor (HFE) to effectively capture the high-frequency components from CBCT images. By wisely integrating high-frequency embedding and CBCT features, this hybrid condition can provide comprehensive guidance that ensures the preservation and reconstruction of fine anatomical structures in the diffusion process, therefore enhancing the quality of the generated images. 
Additionally, we employ the denoising diffusion implicit model (DDIM) during inference, to further speed up the sampling process and dramatically reduce the time for sCT image generation.
We introduce a new in-house prostate dataset and conduct extensive experiments on it, employing widely used image similarity metrics and clinical dosimetric studies in collaboration with our medical physicist team. Our method outperforms state-of-the-art approaches across all evaluation metrics for CBCT-to-CT synthesis. 
The main contributions of this paper are as follow:
\begin{itemize}
    \item We propose a novel hybrid conditional latent diffusion model for CBCT-to-CT synthesis, which leverages both CBCT and its high-frequency image as guidance in diffusion process. We employ UFE to map both images into the latent space and utilize DDIM for generation, to enhance the efficiency of image synthesis.
    \item We design a new HFE to capture high-frequency knowledge from CBCT images. Integrated with CBCT in the latent space, it serves as a hybrid conditional input to the diffusion model, improving sCT image quality by preserving fine anatomical details.
    \item We construct an in-house prostate dataset to facilitate this task, showcasing that HC$^3$L-Diff achieves performance gain in image similarity metrics with reduced inference cost, superior to previous state-of-the-arts.
    \item We study the potential usage of our method in improving clinical dose calculation. Our model outperforms others in dosimetric evaluation metrics, achieving a 93.8\% gamma passing rate (GPR) with a 2\%/2mm criterion. The promising result highlights its great potential for effective treatment planning and adaptive radiotherapy in real-world practice.
\end{itemize}

\section{Methods}
\begin{figure*}
    \centering
    \includegraphics[width=1\linewidth]{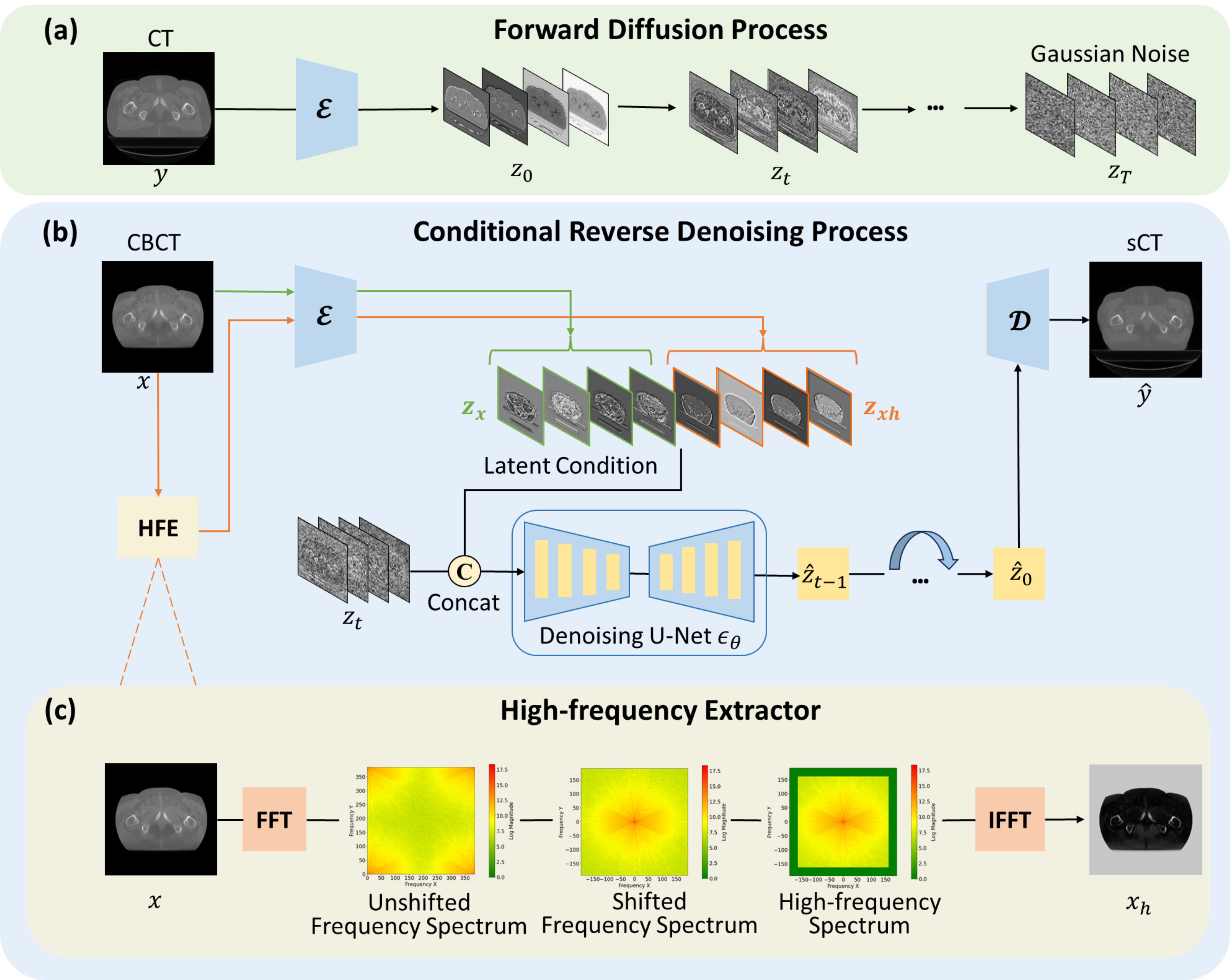}
    \caption{Overview of our proposed HC$^3$L-Diff: (a) In the forward diffusion process, the UFE \(\mathcal{E}\) encodes the CT image \(y\) into latent space representation \(z_{0}\) with 4 channels to realize image compression, and the noise is gradually added to get the standard Gaussian noise \(z_{T}\). (b) During the conditional reverse denoising process, the CBCT image \(x\) and the obtained high-frequency image \(x_{h}\), derived from the HFE, are encoded with \(\mathcal{E}\) respectively to acquire latent features \(z_{x}\) and \(z_{xh}\). These features are integrated in the latent space and serve as the hybrid condition for U-Net, which predicts noise at each time step to facilitate step-by-step denoising until generating \(\hat{z}_{0}\). Finally, the decoder \(\mathcal{D}\) is used to convert it from latent space back into pixel space and get the sCT \(\hat{y}\). (c) The HFE processes CBCT images by applying FFT and FFT shift to preserve high-frequency information, and then utilizes IFFT to get the high-frequency images of CBCT.} 
    \label{fig:framework}
\end{figure*}

In this section, we first introduce the essential information of materials and data characteristics. We then illustrate the framework of the novel hybrid conditional diffusion model, especially on how to integrate and leverage both conditions. Next, we explain how the HFE capture the high-frequency information to refine the generation process and improve the quality of the sCT images. Finally, we describe the objective loss function, training and inference procedure of the model.
The overall framework of our proposed HC$^3$L-Diff is presented in Figure \ref{fig:framework}.

\subsection{Materials and data characteristics}
One key obstacle to this task is the lack of appropriate and publicly available datasets containing paired CBCT and CT for model learning. In this regard, we newly construct a large CBCT-to-CT synthesis database to provide materials in this study. 
Approved by the SingHealth institutional review board, we collect the CBCT and CT data from 100 high-risk prostate cancer treated at National Cancer Centre Singapore (NCCS) between January 2016 and December 2019. Out of the 100 cases, 80 are assigned to train the models and 20 are assigned as the testing set. The CBCT images are acquired using either a Trilogy or iX linear accelerator (LINAC) from the Varian medical system. These are half-fan mode acquisitions with a field-of-view of 45 cm and 120 kVp. The CBCT images have a slice thickness of 2.5 mm and a fixed in-slice resolution of 384 x 384 pixels. The CT data are acquired with a GE LightSpeed RT 16 (GE HealthCare, Chicago, US) with 2.5 mm slice thickness and 120 kVp, which are used as ground truth for model training.

To pair CBCT and CT images one-to-one for generation experiments, we register CT images to match the size of CBCT images using the Plastimatch \cite{sharp2010plastimatch}. For the training set, 80 paired images are divided axially into 5120 CT slices and 5120 CBCT slices. And 1280 paired slices from the remaining 20 patients are used as the testing set. The signal intensity of these images is clipped to [-1000, 3000] and then normalized to [-1, 1] by linear normalization before training and testing.

\subsection{Hybrid conditional latent diffusion}
\noindent\textbf{Preliminaries of diffusion model.} Diffusion models are a type of probabilistic generative model that contains two stages: the forward diffusion process and the reverse generation process \cite{ho2020denoising}. In the forward diffusion process, Gaussian noise is gradually added to the original data until it becomes a random noise. In the reverse generation process, the diffusion model generates sample data by gradually denoising the random noise. The key to the diffusion model is to train a noise prediction model so that the noise predicted by the model for each time step \(t\) is consistent with the real added noise. DDPM adopts a U-Net-like model with residual block as the noise prediction model, and time embedding to indicate time step is injected to each residual block \cite{ho2020denoising}. 
Existing state-of-the-art model for CBCT-to-CT synthesis employs conditional DDPM, which requires multi-step iterations in generation, resulting in high computational complexity, and a long training and inference time \cite{peng2024cbct}.

We propose the unified feature encoder (UFE) as the comprehensive image compression model to transform the high-dimensional data to a lower-dimensional and manageable latent space. The transformation preserves essential information while reducing the overall computational complexity, significantly decreasing the training time and more importantly sample generation time.
Concretely, in the forward diffusion process of model training (Figure \ref{fig:framework} (a)), the UFE encodes the CT image \(y\) into the latent embedding space to realize image compression: \(z_{0}=\mathcal{E}(y)\). We set the downsampling factor \(f = H/h = W/w = 8\), and the channels of the latent embedding as 4, following experience of latent diffusion models \cite{rombach2022high}. 
Subsequently, we incrementally add Gaussian noise to \(z_{0}\) over a series of time steps to produce \(z_{t}\), progressively transforming the latent representation towards standard Gaussian noise \(z_{T}\). 

Next, we propose a novel hybrid condition mechanism in the reverse denoising process of diffusion, as shown in Figure \ref{fig:framework} (b). The hybrid condition consists of the CBCT image \(x\) and its high-frequency information captured by our high-frequency extractor (HFE) from CBCT (more details illustrated in Sec. \ref{sec:highfre}).
During this reverse denoising, we also utilize the UFE to transform both CBCT image \(x\) and its high-frequency image \(x_{h}\) to the latent embedding space: \(z_{x} = \mathcal{E}(x)\), \(z_{x_{h}} = \mathcal{E}(x_{h})\).
We then exploit the concatenation operation to integrate CBCT embedding \(z_{x}\) and the corresponding high-frequency embedding \(z_{x_{h}}\) along the channel dimension into the final hybrid embedding \(\mathcal{C}\) as condition:
\begin{equation}
\mathcal{C}  = [\mathcal{E}(x) \cdot \mathcal{E}(x_{h})]
\end{equation}
where \( \cdot \) denotes the concatenation along the channel dimension. Notably, another advanced method to fuse multiple information is cross-attention. However, we empirically found that the attention technique is inappropriate in our problem setting (experimental comparison shown in Sec. \ref{exp:abl}). On the one hand, cross-attention involves a quadratic complexity due to pairwise interactions, greatly increasing the computational cost and generation time \cite{papanastasiou2023attention}. This presents a major obstacle for its application in online radiotherapy. On the other hand, the massive parameters in cross-attention may increase the optimization difficulty in the diffusion model training process. It turns out that concatenation can yield better results than cross-attention in guiding the iterative sCT generation process.

Conditioned on this hybrid embedding, we then feed the noise-added embedding into the denoiser, a U-Net-like architecture, to predict noise \cite{ho2020denoising}.
At each time step \(t\), we incorporate the hybrid condition \(\mathcal{C}\) into compressed and noise-added CT embedding \(z_{t}\) as the final input, and forward it to the denoiser to predict the corresponding noise. The time embedding is also added to each residual block of the model during this process. 
With predicted noise, we employ the scheme in DDPM to produce the synthesized latent embedding $\hat{z}_{t-1}$. Overall, the reverse denoising process with our novel hybrid condition can be expressed as:
\begin{equation}
\hat{z}_{t-1} = \frac{1}{\sqrt{\alpha_t}} \left( z_t - \frac{1 - \alpha_t}{\sqrt{1 - \bar{\alpha}_t}} \epsilon_\theta(z_t, t, \mathcal{C}) \right) + \sigma_t \mathbf{z}
\label{eq:2}
\end{equation}
\noindent where \(\mathcal{C}\) is our hybrid condition, \(\epsilon_\theta(z_t, t, \mathcal{C})\) is the noise predicted by the denoiser \(\epsilon_{\theta}\) with parameters \(\theta\) given the latent variable \(z_t\), time step \(t\), and condition \(\mathcal{C}\).
\(\alpha_t\) is a constant derived from the variance schedule, controlling the step size, and \(\bar{\alpha}_t\) is the cumulative product of \(\alpha_t\) over all time steps; 
\(\mathbf{z}\) is random noise, sampled from \(\mathcal{N}(0,1)\), with the exception of the time step \(t=1\), where \(\mathbf{z}=0\); \(\sigma_t\) is the noise standard deviation.
Starting from \(t=T\), the aforementioned procedure delineated in Eq. (\ref{eq:2}) is iteratively executed at every time step, ultimately resulting in \(\hat{z}_{0}\).

After the sampling process completes, the trained CT decoder \(\mathcal{D}\) converts the synthesized latent embedding \(\hat{z}_{0}\) back into the pixel space, producing the reconstructed sCT image with the resolution back to the original CBCT: \(\hat{x}=\mathcal{D}(\hat{z}_{0})\) \cite{rombach2022high}. 
The proposed HC$^3$L-Diff framework harmonizes forward and reverse diffusion processes and incorporates a novel hybrid condition in reverse process, efficiently generating clinically relevant sCT images with high quality.

\subsection{Informative high-frequency embedding}
\label{sec:highfre}
High-frequency information is crucial for the CBCT-to-CT synthesis task, as it captures fine structural features such as edges and textures. These details are essential for accurate anatomical representation and reliable image synthesis, leading to clearer and more detailed sCT images \cite{li2023zero}. 
To enhance high-frequency feature extraction from CBCT images, we propose to integrate high-frequency knowledge as an additional condition. Unlike traditional methods that rely on high-pass filters or simple frequency domain filtering, which often involve basic filter designs to remove low-frequency components, our approach leverages both fast Fourier transform (FFT) and FFT shift. The novelty lies in using these techniques to more effectively capture and visualize high-frequency components, leading to richer and more detailed feature extraction.

Concretely, we design a high-frequency extractor (HFE), which adopts the FFT \cite{brigham1988fast} to preprocess the CBCT images, while retaining some high-frequency information. As shown in Figure \ref{fig:framework} (c), the process begins with applying the FFT and FFT shift operations to the CBCT image \(x\):
\begin{equation}
X_{f} = \mathcal{F}_{\text{shift}}\left( \mathcal{F}\left( x \right) \right)
\end{equation}
where $\mathcal{F}$ denotes the FFT, and $\mathcal{F}_{\text{shift}}$ represents the FFT shift operation. The FFT converts the CBCT image from the spatial domain to the frequency domain, allowing for the analysis of different frequency components within the image. The FFT shift operation then moves the zero-frequency component to the center of the frequency spectrum. This central positioning of the zero-frequency component enhances the visualization and analysis of both low and high-frequency components, making it easier to interpret the frequency distribution and identify significant features across the spectrum.

Then, high-pass filtering is achieved by setting the low-frequency part to zero, preserving the high-frequency information in the image. The cutoff frequency \(th\) is empirically set to 30, where \(f_x\) and \(f_y\) represent the horizontal and vertical frequency components, respectively:
\begin{equation}
X_{h} = \begin{cases} 0, & \text{if } |f_x| < {th} \text{ or } |f_y| < {th} \\ X_{f}, & \text{otherwise} \end{cases} 
\end{equation}
Finally, the inverse FFT shift and inverse FFT (IFFT) are consecutively performed to transform the high-frequency component back to the spatial domain:
\begin{equation}
x_{h} = \mathcal{F}^{-1}\left( \mathcal{F}^{-1}_{\text{shift}}\left( X_{h} \right) \right)
\end{equation}
where $\mathcal{F}^{-1}$ denotes the IFFT, and $\mathcal{F}^{-1}_{\text{shift}}$ represents the inverse FFT shift operation.
The obtained $x_{h}$ is injected into the diffusion model as the condition to provide informative high-frequency knowledge to benefit sCT generation with more structural details recovered.

\subsection{Objective function and learning process}
The training of our HC$^3$L-Diff consists of two stages:
\textit{1) Training UFE and decoder:} we first train the UFE and decoder by the image reconstruction task, using the architecture of vector quantized variational autoencoder \cite{kingma2013auto}. We train them by a combination of perceptual loss, structural similarity index (SSIM) loss, quantization loss, and L1-based reconstruction loss, following previous literature on image reconstruction \cite{esser2021taming, wang2004image, van2017neural}. Once the UFE and decoder are trained, the parameters of \(\mathcal{E}\) and \(\mathcal{D}\) are frozen for the next stage.
\textit{2) Training conditional latent diffusion model:} we then train the conditional latent diffusion model using the pre-trained UFE \(\mathcal{E}\). 
The total number of time steps \(T\) is set to 1000. For time step \(t\), given the latent representation \(z_{t}\) of the CT image and the hybrid condition \(\mathcal{C}\) as the input, we train the conditional latent diffusion model by computing the L1 loss:
\begin{equation}
L=E_{\mathcal{E}(x),\epsilon\sim \mathcal{N}(0,1), t}\left \| \epsilon - \epsilon_{\theta } (z_{t},t,\mathcal{C}) \right \|  
\end{equation}
where \( \epsilon \) is the Gaussian noise sampled from \(\mathcal{N}(0,1)\); \( \epsilon_{\theta} (z_{t}, t, \mathcal{C}) \) denotes the predicted noise at the time step \( t \) by our hybrid conditional denoiser; since the forward process is fixed, \( z_{t} \) can often be derived from $\mathcal{E}(x)$ by adding noise according to a predefined schedule over the course of several steps from \( t=0 \) to \( t=T \).

During the inference stage, we use DDIM to accelerate the sampling process and significantly shorten the image generation time \cite{song2020denoising}. The DDPM generation process is a reverse Markov chain, requiring numerous steps (often around 1000) to generate images, which is slow and inefficient for clinical applications \cite{ho2020denoising}. In contrast, DDIM shares the same training process with the DDPM, but employs a non-Markovian process, the reverse process is modified so that each step can reverse multiple forward diffusion steps. 
Its sampling process can be expressed as follow:
\begingroup
\small
\begin{equation}
    \hspace{-1em} % Adjust the value to move the equation left
    z_{t-1} = \sqrt{\alpha_{t-1}} \left( \frac{z_t - \sqrt{1 - \alpha_t} \epsilon_{\theta} (z_t, t, \mathcal{C})}{\sqrt{\alpha_t}} \right) + \sqrt{1 - \alpha_{t-1}} \epsilon_{\theta} (z_t, t, \mathcal{C})
\end{equation}
\endgroup
where \(\alpha_t\) is a constant derived from the variance schedule, and it controls the rate of diffusion and influences the sampling efficiency.
We exploit DDIM in inference to produce high-quality samples with substantially fewer steps, achieving faster execution times and therefore further enhancing the practical applicability of the model to online adaptive radiotherapy. 

\section{Experiments}
We validate the effectiveness of our CBCT-to-CT synthesis approach on our in-house prostate dataset, with method comparison via both qualitative and quantitative analysis. We perform extensive ablation analysis and also study the potential usage of our method in improving real clinical dose calculation.

\subsection{Implementation details}
We train both the UFE and the conditional latent diffusion model on a 40GB NVIDIA A100 GPU. The UFE requires 200 epochs for training, while the conditional latent diffusion model is trained for 1000 epochs. The diffusion model utilizes a Gaussian noise scheduler with 1000 time steps and a linearly scaled noise level from 0.002 to 0.02, and the AdamW optimizer with a learning rate of 1e-4 is used. The sampling time step of DDIM is set to 150. Besides, all the comparison methods are re-implemented using their released code, default configuration, and parameters provided in the original papers for fair comparison.

\subsection{Evaluation metrics}
We select the mean absolute error (MAE), peak signal-to-noise ratio (PSNR), and SSIM as the evaluation metrics of the image generation effect \cite{deng2023synthetic}. The specific calculation formulas are as follows:
\begin{equation}
\text{MAE}=\frac{1}{n}\sum_{i=1}^{n}\left | CT_{i}-sCT_{i}  \right |
\end{equation}

\begin{equation}
\text{PSNR}=10log_{10}\frac{\text{Max}^{2}}{\frac{1}{n}\sum_{i=1}^{n}\left ( CT_{i}-sCT_{i} \right )^{2}}
\end{equation}

\begin{equation}
\text{SSIM}=\frac{(2\mu _{CT}\times \mu _{sCT}+  C_{1}  )(2\sigma_{CTsCT}  +C_{2} )}{(\mu _{CT}^{2} +\mu _{sCT}^{2} +  C_{1})(\sigma_{CT}^{2} +\sigma_{sCT}^{2} +C_{2} )} 
\end{equation}
where \(n\) is the total number of pixels in the images, \(CT_{i}\) and \(sCT_{i}\) are the pixel values in the CT and sCT images, respectively. Max is set to 3000 HU, representing the maximum pixel intensity in CT images. \(\mu_{CT}\) and \(\mu_{sCT}\) are the mean pixel values, \(\sigma_{CT}\) and \(\sigma_{sCT}\) are the standard deviations, and \(\sigma_{CTsCT}\) is the covariance of the CT and sCT images. \(C1 = (0.01\text{Max})^2\) and \(C2 = (0.03\text{Max})^2\) are constants used to stabilize the division by weak denominators.

\subsection{Comparison with state-of-the-art methods}

\begin{table*}
\centering
\caption{Quantitative experiment results of different sCT images on the prostate dataset.}\label{tab1}
% \begin{tabular}{|l|l|l|}
\begin{tabular}{ccccc}
% \begin{tabular}{|c|c|c|}
\toprule
 Method &  MAE \(\downarrow\) & PSNR \(\uparrow\) & SSIM \(\uparrow\) & Time \(\downarrow\) \\
 \midrule
%CBCT &  70.966 &{25.779} & 0.741 &  / \\
CycleGAN \cite{zhu2017unpaired} & 151.636 & 20.193 & 0.608 & 2s \\
MaskGAN \cite{phan2023structure} & 88.350  &23.609 & 0.679 & 556s  \\
Conditional DDPM \cite{peng2024cbct} & 96.919  & 25.305 &0.730   & 56min  \\
\textbf{HC$^3$L-Diff (Ours)}  & \textbf{53.614} &  \textbf{26.356} & \textbf{0.802}  & \textbf{142s} \\
\bottomrule
\end{tabular}
\label{tb:comparison}
\end{table*}

To assess the efficacy of our proposed model, we conducted comprehensive comparison experiments against several state-of-the-art generative models based on GAN or diffusion models, including CycleGAN~\cite{zhu2017unpaired}, MaskGAN~\cite{phan2023structure}, and conditional DDPM~\cite{peng2024cbct}. The quantitative experiment results of different methods are presented in Table \ref{tab1}. Evaluation results across various models indicate that the proposed HC$^3$L-Diff achieves superior CBCT-to-CT image generation with high quality in only 142 seconds, which meets the clinical requirement in time (less than 5 minutes). This performance significantly outperforms other methods. Despite CycleGAN's shorter generation time, the quality of its sCT is substantially inferior to HC$^3$L-Diff, failing to meet the standards required for clinical applications. Although MaskGAN achieves improvement over CycleGAN, its output still falls short in image quality and suffers from long-time processing issues. Conditional DDPM showcases the potential advantages of diffusion models, achieving better PSNR, and SSIM scores than GAN-based approaches. However, its extensive computational demands, requiring 56 minutes to process a single 3D image, render it impractical for clinical applications.

In contrast, HC$^3$L-Diff achieves an optimal balance between image quality and computational efficiency, positioning it as a promising solution for clinical CBCT-to-CT synthesis. The integration of UFE and DDIM significantly reduces computational time, enabling our model to generate high-quality sCT images in only 142 seconds per patient, which is around $\times$20 times faster than conditional DDPM. UFE efficiently encodes images into a compact latent space, while DDIM accelerates the diffusion process without compromising image quality. Regarding image quality, HC$^3$L-Diff outperforms conditional DDPM because the integration of the HFE and UFE enables the model to more effectively capture intricate anatomical structures and generate highly realistic images through an optimized reverse denoising process. The HFE focuses on preserving fine details, while the UFE ensures comprehensive feature representation, collectively contributing to the generation of realistic sCT images in reverse diffusion denoising. 

\begin{figure*}
    \centering
    \includegraphics[width=1\linewidth]{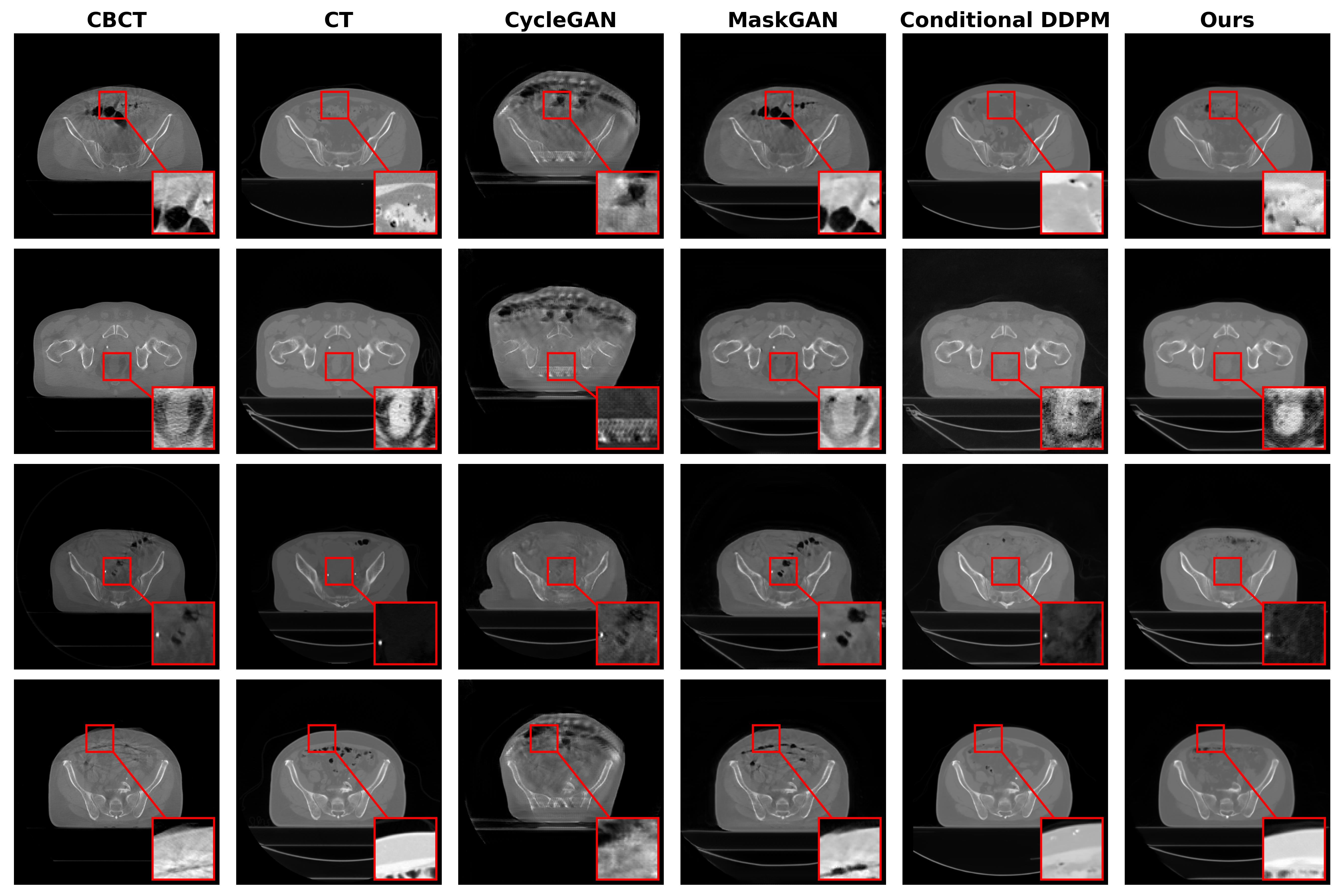}
    \caption{Qualitative comparison of different generative methods, and the four rows represent prostate images from four different patients.}
    \label{fig:comparison}
\end{figure*}

Figure \ref{fig:comparison} presents a visual comparison of different methods. We visualize the original CBCT images, CT images (ground truth), and the sCT generation results of the four typical test cases. To more clearly demonstrate the results of the different methods, we have also zoomed in on local regions within the images. 
We can see that the sCT images generated by CycleGAN exhibit the lowest quality, showcasing numerous artifacts (as shown in the first and fourth row) and unrealistic anatomical structures (as shown in the second and third row), along with distracting noise and incomplete regions (as shown in the first row), thus making them even inferior to the original CBCT images. While MaskGAN and conditional DDPM maintain the general organ outline comparable to the original CBCT, they still fall short in accurately replicating specific details compared to the real CT images.  
For the incomplete regions in CBCT, as seen in the zoomed-in areas in the first and third row, the sCTs generated by MaskGAN are still incomplete.  
Regarding the noises and artifacts in CBCT, as depicted in the zoomed-in areas in the second and fourth rows, the MaskGAN fails to remove them effectively, and the sCT generated by conditional DDPM is still blurry (second) or fails to reconstruct correct anatomical structure (fourth).
In contrast, our method not only preserves the overall anatomical structure with high quality but also accurately reproduces fine details that are crucial for clinical diagnosis, as highlighted in the zoomed-in regions.

\begin{figure*}
    \centering
    \includegraphics[width=0.8\linewidth]{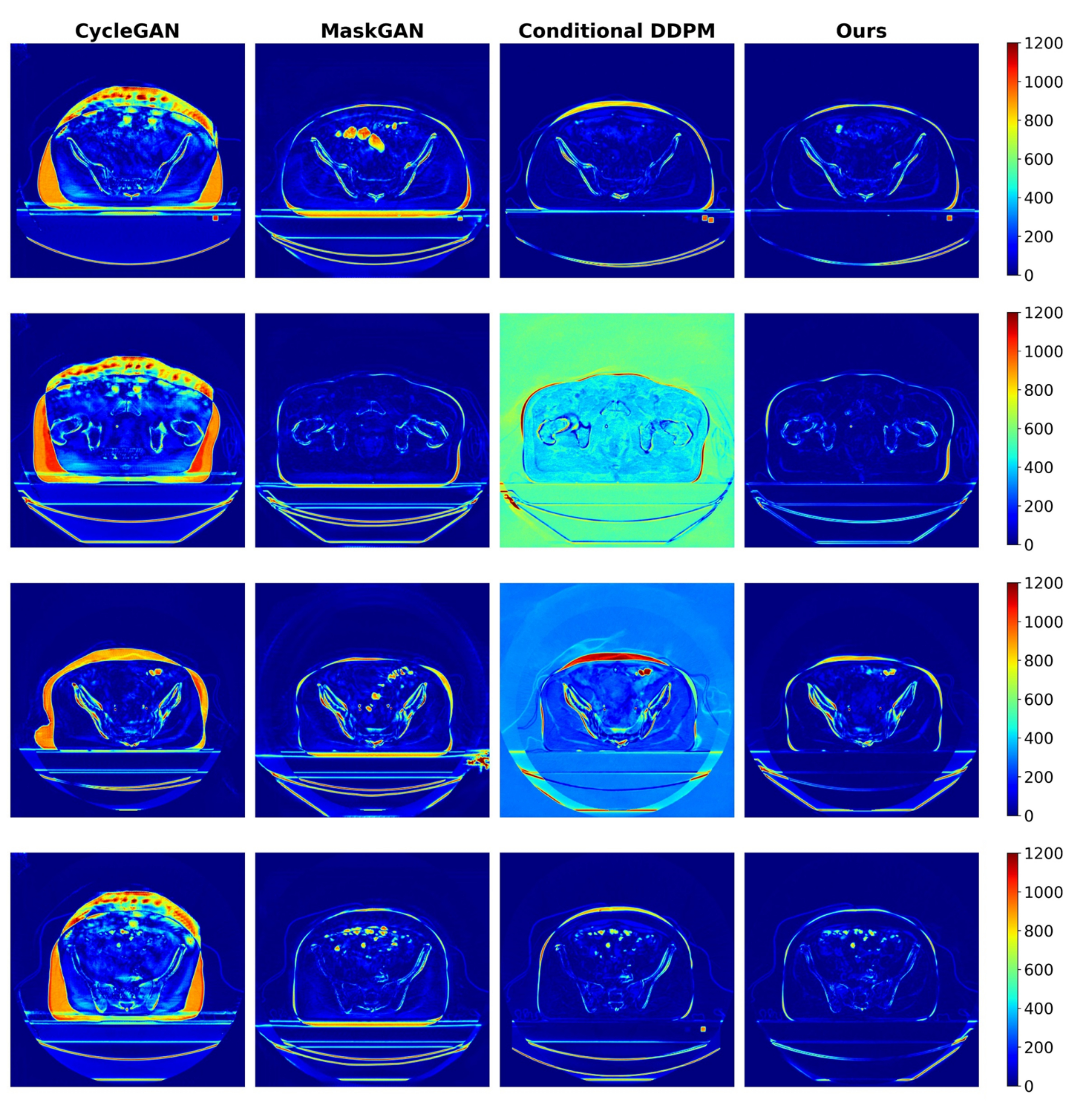}
    \caption{HU value difference maps of CT images and sCT images generated by different methods. Each row corresponds to the results from one patient.}
    \label{fig:hu_difference}
\end{figure*}

To further compare the generative performance of different methods, we illustrate the difference in Hounsfield units (HU) value between CT and sCT images generated by different methods in Figure \ref{fig:hu_difference}, with a higher value representing lower performance. HU values, a measure of radiodensity used in computed tomography, are crucial for precise dose calculations in radiotherapy. Therefore, comparing HU value differences is significant in this study to evaluate the accuracy and clinical relevance of the generated images. Each row represents the results for each patient.
We can observe that a substantial region exhibits a pronounced HU value disparity between the sCT images generated by CycleGAN and the corresponding CT images, signifying that CycleGAN encounters significant challenges in accurately rendering complex anatomical structures.
Although the MaskGAN-generated sCT images show smaller differences than those of CycleGAN, significant discrepancies persist in organ boundary regions, such as the lower part of the organ in the first and fourth row. Moreover, MaskGAN also exhibits generation biases within the internal areas of organs, as seen in the first and third row. This indicates that although MaskGAN surpasses CycleGAN in performance, it still fails to accurately reconstruct the fine anatomical and boundary details crucial for accurate calculation of dose distribution. 
Conditional DDPM also shows suboptimal performance in organ contour areas, as observed on the upper side of the organ in the first and third row. For the challenging sample, the entire image generated by conditional DDPM may have a large difference, as shown in the second row.
Compared to the other models, our HC$^3$L-Diff model demonstrates clear advantages, manifesting better generative performance in rendering both intricate internal organ details and delineating contour regions. 
Our HC$^3$L-Diff method achieves superior performance in both internal organ details and contour areas. This superiority arises from the integration of high-frequency image information, which enhances the model's ability to capture detailed structural features and sharp edges. 
Consequently, our model consistently delivers the most accurate sCT images, maintaining structural integrity for precise dose calculations, making it the most optimal among the compared models.

\subsection{Ablation study}
\label{exp:abl}
\begin{table*}
\centering
\caption{Effectiveness of different components of our proposed model.}\label{tab2}
% \begin{tabular}{|l|l|l|}
\begin{tabular}{c|ccc|cccc}
% \begin{tabular}{|c|c|c|}
\toprule
 % \hline
Method & UFE & DDIM & HFE &  MAE \(\downarrow\) & PSNR \(\uparrow\) & SSIM \(\uparrow\) & Time \(\downarrow\) \\
 % \midrule
 \hline
M1 &  & & & 96.919  & 25.305 &0.730   & 56min \\
M2 & \checkmark  & & & 54.986   & 26.187 & 0.798 & 625s  \\
M3 & \checkmark & \checkmark  &  & 54.508 & 26.251 & 0.799 & 130s \\

 \hline
Ours & \checkmark & \checkmark  & \checkmark &\textbf{53.614} &  \textbf{26.356} & \textbf{0.802}  & \textbf{142s} \\
 \bottomrule
\end{tabular}
\label{tb:ablation}
\end{table*}

We conduct ablation experiments to validate the effectiveness of different key components in the proposed method and obtain four configurations. The results are shown in Table \ref{tab2}. M1 represents the baseline, conditional DDPM, which takes 56 minutes to generate a 3D volume of a patient. M2, which integrates UFE as an image compression model based on the baseline, can generate a 3D volume in 625s, reducing 80\% time compared to the baseline. Additionally, the MAE of the generated image is significantly reduced, and both PSNR and SSIM are improved. This demonstrates that the inclusion of UFE not only substantially improves the speed of sCT image generation but also effectively improves the quality of the generated images. 
M3 further replaces DDPM with DDIM in the reverse denoising process based on M2. This improves inference speed and enhances the quality of the generated images, underscoring the superiority of DDIM in inference acceleration. 
By wisely incorporating high-frequency knowledge as condition, our complete model further increases the quality of sCT with little inference time gain, particularly reducing the MAE by 0.9. The results can reflect intricate structural information in more detail, which is critically important for accurate dose calculation and adaptive radiotherapy.

\begin{table*}
\centering
\caption{Results of different ways to integrate time embeddings and conditions in diffusion model.}\label{tab3}
% \begin{tabular}{|l|l|l|}
\begin{tabular}{c|cc|cc|cccc}
% \begin{tabular}{|c|c|c|}
\toprule
 % \hline
Method & Add \(t\) & Att with \(t\) & Concat condition & Att with condition &  MAE \(\downarrow\) & PSNR \(\uparrow\) & SSIM \(\uparrow\) & Time \(\downarrow\) \\
 % \midrule
 \hline
C1 &   & \checkmark & & \checkmark  &56.466
 & 26.042 & 0.790 & 418s \\
C2 &   & \checkmark & \checkmark  &  & 54.184& 26.288 & 0.799 &390s \\
C3 & \checkmark &  &   & \checkmark & 54.877 & 26.238  & 0.796 & 201s \\
\hline
Ours & \checkmark & & \checkmark  & &  \textbf{53.614} &  \textbf{26.356} & \textbf{0.802}  & \textbf{142s} \\
 \bottomrule
\end{tabular}
\label{tb:combination}
\end{table*}

How to integrate condition information and time embeddings plays a crucial role in accurate CBCT-to-CT generation. We then conduct experiments with different integration settings and list the results in Table \ref{tab3}. Regarding adding time embeddings to the noise map, two settings are compared: element-wise addition and cross-attention fusion. We can see that the addition yields superior image generation results. 
Regarding the fusion of condition information derived from CBCT and high-frequency images, we compared two settings: concatenating conditions in the channel dimension and employing cross-attention with conditions. We can see that concatenating conditions attains better results. Overall, we propose to use addition operations to integrate time information and utilize concatenation to introduce conditions into the model to predict noise.

\subsection{Clinical validation with dosimetric study}

\begin{figure*}
    \centering
    \includegraphics[width=0.9\linewidth]{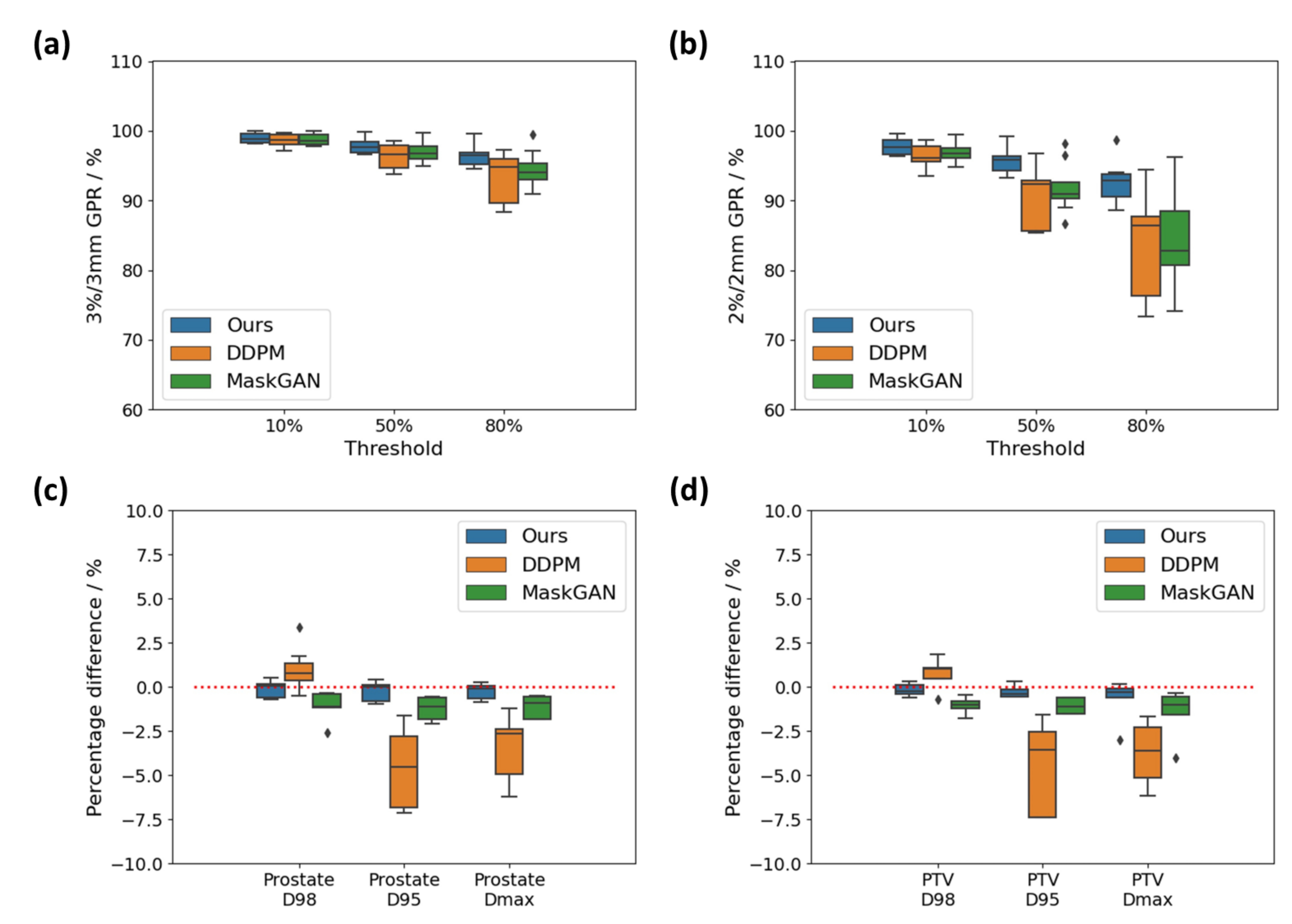}
    \caption{Figures (a) and (b) show the boxplots of the GPRs between the sCT and the ground truth CT for three different dose thresholds using the 3\%/3mm and 2\%/2mm gamma criteria respectively. Figures (c) and (d) show the percentage difference in the DVH parameters (D95, D98, and Dmax) between the sCT and the ground truth CT for prostate and PTV respectively.}
    \label{fig:gpr}
\end{figure*}

\subsubsection{Dosimetric evaluation metrics}
Apart from image similarity metrics, the model can also be assessed based on the dose agreement calculated in the generated sCT and a ground truth CT. The ground truth is created from deformable image registration (DIR) from the planning CT to the CBCT. To qualify as a reliable ground truth CT, the anatomies must be mapped accurately so that there should be minimal anatomical difference between the ground truth CT and the CBCT. It is well known that DIR often fails when there are large anatomical differences between the planning CT and the CBCT \cite{dir_1, dir_2}, and hence, the ground truth CTs need to be carefully curated. In this study, our medical physicist (H. Q. Tan) curated the test dataset to select 10 (out of the 20) patients with reliable ground truth CT. 
The doses are calculated on the ground truth CT and the sCT generated by different methods, using the Monte Carlo algorithm in the RayStation 2023B (Raysearch Laboratories, Stockholm, Sweden) treatment planning system (TPS). The patients were all treated with 10 MV photon beams using a volumetric arc therapy (VMAT) technique. 

Two evaluation methods on dose comparison are used in this work. The first is the gamma analysis approach which is widely used in the radiation oncology field \cite{low2003}. Loosely speaking, this approach assesses the dose agreement between two dose distributions (in this case, dose calculated on ground truth CT and sCT). It has the advantage of including both dose difference and distance-to-agreement, which makes it a reliable metric for assessing dose difference in the high dose gradient area of the treatment plan \cite{depuydt2002}. It is defined as:

\begingroup
\small
\begin{equation} 
\hspace{-1em} % Adjust the value to move the equation left
\Gamma (\vec{r}_{ref}, \vec{r}_e) = \sqrt{\frac{|\vec{r}_{ref}-\vec{r}_e|^2}{\delta r^2} + \frac{[D_e(\vec{r}_e) - D_{ref}(\vec{r}_{ref})]^2}{\delta D^2} } 
\end{equation}
\endgroup

\noindent where \(\vec{r}_e\) and \(\vec{r}_{ref}\) represent the positions of the voxels of the evaluation and reference doses respectively. \(D(\vec{r})\) is the dose at voxel position \(\vec{r}\). \( \delta D \) and \( \delta r \) are the respective dose and distance criteria. \( \delta D \) is usually quoted as a percentage of the maximum dose in the reference dose distribution. The metric used in gamma analysis is the gamma passing rate (GPR) which represents the proportion of voxels with \( \Gamma < 1 \). In this study, the GPR is calculated for 3\%/3mm and 2\%/2mm criteria and for three dose thresholds namely, 10\%, 50\%, and 80\% of the maximum dose. Only voxels with doses greater than the threshold are included in the GPR computation; this allows us to focus on dose agreement in different dose regions of the treatment plan. A higher GPR represents better dose agreement between the two dose distributions calculated on ground truth and sCT, and hence better generated sCT.

The second method of dose evaluation is based on the agreement between the dose volume histogram (DVH) clinical goals. D98, D95, and Dmax are calculated for both the prostate and planning target volume (PTV) for the ground truth CT and sCTs. The percentage differences between the DVH parameters of the sCTs and ground truth CT are used to quantify the degree of dose agreement. An absolute value closer to zero signifies a diminished discrepancy from the ground truth, thereby denoting superior generation performance.

\subsubsection{Dosimetric evaluation of the model}
The results of the dosimetric comparison between different methods are shown in Figure \ref{fig:gpr}. CycleGAN is omitted from further dosimetric analysis as most of its sCTs are distorted and unintelligible (refer to Figure \ref{fig:comparison}) which makes it unsuitable for clinical use. Figure \ref{fig:gpr} (a) and (b) show the GPR comparison for 3\%/3mm and 2\%/2mm respectively. Our proposed method achieves the highest GPR among the three models across all the dose thresholds. Notably, all the GPRs are greater than 90\% for our proposed model, indicating that most of the voxels have low dose errors of less than 2\%. However, DDPM and MaskGAN models yield GPRs of less than 90\% with 2\%/2mm criterion, especially within the target at the 80\% dose threshold. 

The DVH comparisons are shown in Figure \ref{fig:gpr} (c) and (d) for the prostate and PTV respectively. Similarly, the proposed model yields the smallest percentage difference in D95, D98, and Dmax with the ground truth CT. In particular, the PTV D95 dose differences are less than 1.0\% for the proposed model, which shows that despite the dose differences shown in the GPR, there is negligible impact on the clinical conclusion. 

\begin{figure*}[th!]
    \centering
    \includegraphics[width=0.9\linewidth]{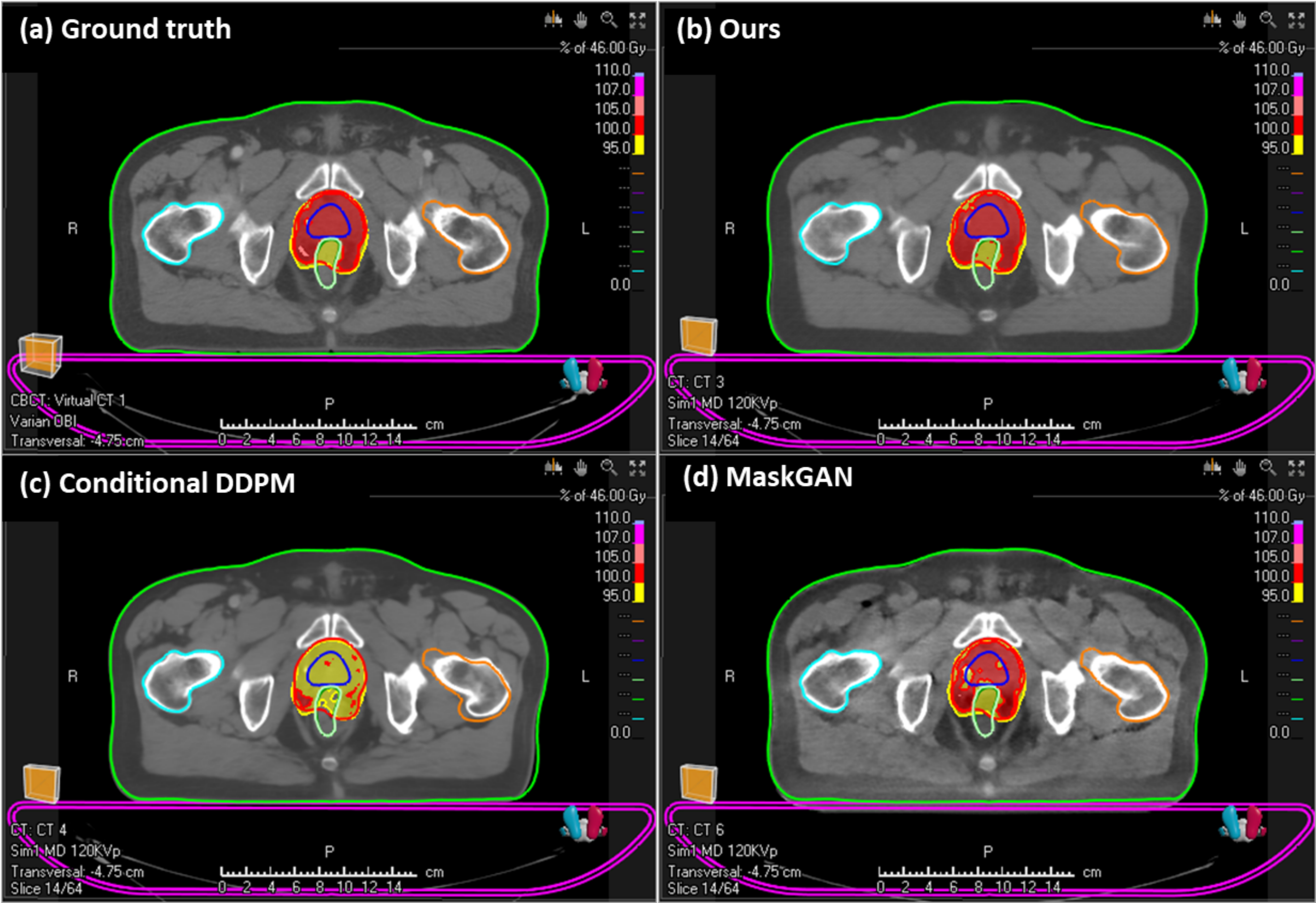}
    \caption{Figure of the dose distribution in the ground truth CT and the three different sCT algorithms from a VMAT delivery. The yellow and red color wash shows the 95\% and 100\% isodose regions. The prostate and the PTV are shown by the blue and red curves respectively.}
    \label{fig:dose}
\end{figure*}

Figure \ref{fig:dose} shows the dose distribution calculated in the TPS based on ground truth CT and the sCTs. This represents the nominal patient result in the test dataset. DDPM and MaskGAN models have a 2\%/2mm GPRs of 86.3\% and 86.7\% respectively with an 80\% dose threshold while the proposed model has a GPR of 93.8\%. Despite this difference, it is apparent that the 95\% 
isodose region, indicated by the yellow heatmap are largely similar in all the figures. In fact, both Figure \ref{fig:dose} (b) and (d) show the nice exclusion of the rectum by the 100\% isodose region which exists in the the ground truth result (Figure \ref{fig:dose} (a)) as well. DDPM shows the largest difference from the rest of the models where the 100\% isodose region (indicated by the red heatmap) does not cover the prostate and PTV at all. 

\section{Discussion}
Our proposed model demonstrates exceptional performance in generating sCT images from 2D CBCT inputs. Leveraging the UFE and DDIM, it achieves efficient CBCT-to-CT image synthesis, and the integration guidance of CBCT image and high-frequency information for the conditional diffusion model enhances the accuracy of our method. It achieves the state-of-the-art results on a real-world prostate dataset. The ensuing dosimetric evaluations also show that the proposed model achieves improved dose agreement with the ground truth CT both in a voxel-wise manner (using GPR) and also from the DVH parameters, especially with regards to the target coverage (D95 and D98 of the prostate and PTV).  

We analyzed the parameter selection of our model, emphasizing both efficiency and quality. Utilizing the DDIM expedites the generation of sCT images, with the number of sampling steps serving as a crucial hyperparameter affecting generation time. Although our DDPM is trained with 1000 steps, we opt for 150 steps for DDIM to strike a balance between speed and image quality. The inference time for a 3D volume with 150 steps is about 142 seconds. Reducing the steps to 100 decreases the time to around 101 seconds but results in a decline in image quality. Conversely, increasing the steps to 200 prolongs the inference time to 174 seconds without any significant improvement in image quality.

Furthermore, as the proposed HC$^3$L-Diff is a data-driven approach and utilizes almost no domain-specific knowledge of prostate data, it is general for various 3D medical image synthesis tasks. While this study focuses on the prostate dataset, our approach is extendable to other organ datasets, such as brain and neck. This suggests our model's versatility and potential for broader clinical application, enhancing its practical value.

Despite the advancements achieved, our study presents limitations and we plan to improve in future research. 
To ease the optimization difficulty in model training, we use 2D convolution as the building block of the proposed model. However, the volumetric information of CBCT and CT has not been sufficiently leveraged. In future, we plan to update the convolution to the 3D version, and meanwhile, propose strategies to alleviate the learning difficulty.
In addition, we develop our method for CBCT-to-CT generation task, handling the single modality. In future, we plan to include multi-modality learning to improve the generalization of proposed method to tackle wider generation applications, such as MRI to CT conversion \cite{boulanger2021deep, wang2022development}, and multi-modality MRI synthesis \cite{dai2020multimodal}, thereby broadening the scope of intelligent healthcare solutions and improving patient care.

\section{Conclusion}
CBCT is crucial for guiding radiation therapy, but its utility is limited by incomplete information and noise, making it unsuitable for precise dose calculations. To address this, we propose a novel hybrid conditional latent diffusion model adapted for CBCT data to synthesize high-fidelity CT images. Our model significantly reduces inference time, generating a 3D CT image in only over two minutes. This efficiency is achieved by the integration of UFE and DDIM for sampling. Including high-frequency information as condition further enhances image quality, allowing our model to replicate fine structural features and sharp edges. Given the hybrid condition of CBCT and high-frequency image, our diffusion model, with its forward and reverse denoising processes, can capture data distributions more effectively and produce more realistic images compared to state-of-the-art models. We establish an in-house prostate dataset for experimental evaluation and conduct comprehensive dose calculations to assess the quality of the synthesized CT images. Experimental results show that our model achieves state-of-the-art performance, with superior MAE, PSNR, and SSIM metrics. Ablation studies confirm the efficacy of each component of our model, and dose metric evaluations demonstrate that our model’s ability to minimize differences between sCT and CT images translates to more reliable and accurate dose distributions. This significantly enhances the effectiveness and safety of radiotherapy treatments.

% Add the CRediT authorship contribution statement section
\section*{CRediT authorship contribution statement}
\textbf{Shi Yin:} Investigation, Conceptualization, Methodology,  Data curation, Software, Validation, Visualization, Writing – original draft. 
\textbf{Hongqi Tan:} Data curation, Investigation, Methodology, Validation, Visualization, Writing - Original draft. 
\textbf{Li Ming Chong:} Data curation, Writing – review \& editing. 
\textbf{Haofeng Liu:} Methodology, Writing – review \& editing. 
\textbf{Hui Liu:} Methodology, Writing – review \& editing. 
\textbf{Kang Hao Lee:} Validation, Writing – review \& editing. 
\textbf{Jeffrey Kit Loong Tuan:} Data curation, Writing – review \& editing. 
\textbf{Dean Ho:} Resources, Funding acquisition, Writing – review \& editing. 
\textbf{Yueming Jin:} Investigation, Project administration, Supervision, Funding acquisition, Writing – review \& editing.

% Add the Declaration of competing interest section
\section*{Declaration of competing interest}
Dean Ho is one of the inventors of previously filed pending patents on artificial intelligence-based therapy development. Dean Ho is one of the co-founders and shareholders of KYAN Therapeutics, which is commercializing intellectual property pertaining to Al-based personalized medicine. The findings from this study are being made available for public benefit, and no intellectual property rights arising from the work reported here are being pursued. The remaining authors declare no conflicts of interest.
%The authors declare that they have no known competing financial interests or personal relationships that could have appeared to influence the work reported in this paper.

\section*{Acknowledgments}
This work was supported by Ministry of Education Tier 1 Start up grant, NUS, Singapore (A-8001267-01-00); Ministry of Education Tier 1 grant, NUS, Singapore (A-8001946-00-00).

\bibliographystyle{cas-model2-names}

\bibliography{cas-refs.bib}

\bio{}
\endbio

\endbio

\end{document}